\documentclass[letterpaper]{article} % DO NOT CHANGE THIS
\usepackage{aaai25}  % DO NOT CHANGE THIS
\usepackage{times}  % DO NOT CHANGE THIS
\usepackage{helvet}  % DO NOT CHANGE THIS
\usepackage{courier}  % DO NOT CHANGE THIS
\usepackage[hyphens]{url}  % DO NOT CHANGE THIS
\usepackage{graphicx} % DO NOT CHANGE THIS
\usepackage{natbib}  % DO NOT CHANGE THIS AND DO NOT ADD ANY OPTIONS TO IT
\usepackage{caption} % DO NOT CHANGE THIS AND DO NOT ADD ANY OPTIONS TO IT
\usepackage{algorithm}
\usepackage{algorithmic}

\usepackage{subfigure}
\usepackage{amsfonts,amssymb}
\usepackage{amsmath}
\usepackage{algorithm}
\usepackage{algorithmic}
\usepackage{bbding}
\usepackage{multirow}

\newcommand{\ie}{\textit{i}.\textit{e}.}
\usepackage{booktabs}
\newtheorem{definition}{Definition}
\newtheorem{proposition}{Proposition}

\usepackage{newfloat}
\usepackage{listings}
\DeclareCaptionStyle{ruled}{labelfont=normalfont,labelsep=colon,strut=off} % DO NOT CHANGE THIS
\floatstyle{ruled}
\newfloat{listing}{tb}{lst}{}
\floatname{listing}{Listing}
\title{Improving Integrated Gradient-based Transferable Adversarial Examples\\by Refining the Integration Path }
\author{
    Yuchen Ren\textsuperscript{\rm 1}, Zhengyu Zhao\textsuperscript{\rm 1}\thanks{Corresponding Author}, Chenhao Lin\textsuperscript{\rm 1}, Bo Yang\textsuperscript{\rm 2}, Lu Zhou\textsuperscript{\rm 3}, Zhe Liu\textsuperscript{\rm 4}, Chao Shen\textsuperscript{\rm 1}\\
  }

\affiliations{
    \textsuperscript{\rm 1}Xi'an Jiaotong University \\
    \textsuperscript{\rm 2}Information Engineering University, 
    \textsuperscript{\rm 3}Nanjing University of Aeronautics and Astronautics, 
    \textsuperscript{\rm 4}Zhejiang Lab\\
    ryc98@stu.xjtu.edu.cn, 
    \{zhengyu.zhao, linchenhao\}@xjtu.edu.cn,\\
    yangbo\_hn@163.com, 
    \{lu.zhou, zhe.liu\}@nuaa.edu.cn, chaoshen@mail.xjtu.edu.cn 
}

\begin{document}

\maketitle

\begin{abstract}

Transferable adversarial examples are known to cause threats in practical, black-box attack scenarios. A notable approach to improving transferability is using integrated gradients (IG), originally developed for model interpretability. In this paper, we find that existing IG-based attacks have limited transferability due to their naive adoption of IG in model interpretability. To address this limitation, we focus on the IG integration path and refine it in three aspects: multiplicity, monotonicity, and diversity, supported by theoretical analyses. We propose the Multiple Monotonic Diversified Integrated Gradients (MuMoDIG) attack, which can generate highly transferable adversarial examples on different CNN and ViT models and defenses. Experiments validate that MuMoDIG outperforms the latest IG-based attack by up to 37.3\% and other state-of-the-art attacks by 8.4\%. In general, our study reveals that migrating established techniques to improve transferability may require non-trivial efforts. 
Code is available at \url{https://github.com/RYC-98/MuMoDIG}.
%% following link form is necessary for aaai2025
%  \link{Appx. \& Code}{https://github.com/RYC-98/MuMoDIG}
\end{abstract}


\section{Introduction}
Deep learning networks (DNNs) are known to be vulnerable to adversarial attacks, which slightly alter input examples to cause model prediction errors \cite{intriguing,fgsm}.
Adversarial attacks can be divided into white-box and black-box attacks. White-box attacks assume that all information about the target model is transparent. Conversely, black-box attacks \cite{vt,ba,epca} are more challenging due to the lack of model details. In particular, transferable black-box attacks \cite{b9} even assume no access to the target model's output, and they just rely on the transferability of adversarial examples generated on known surrogate models.
To assess and then improve the robustness of DNN models in practical scenarios, various transferable attacks have been proposed \cite{ssa,sgm,logit}.

\begin{figure}[!t]
\centering
    \includegraphics[width=1.0\columnwidth]{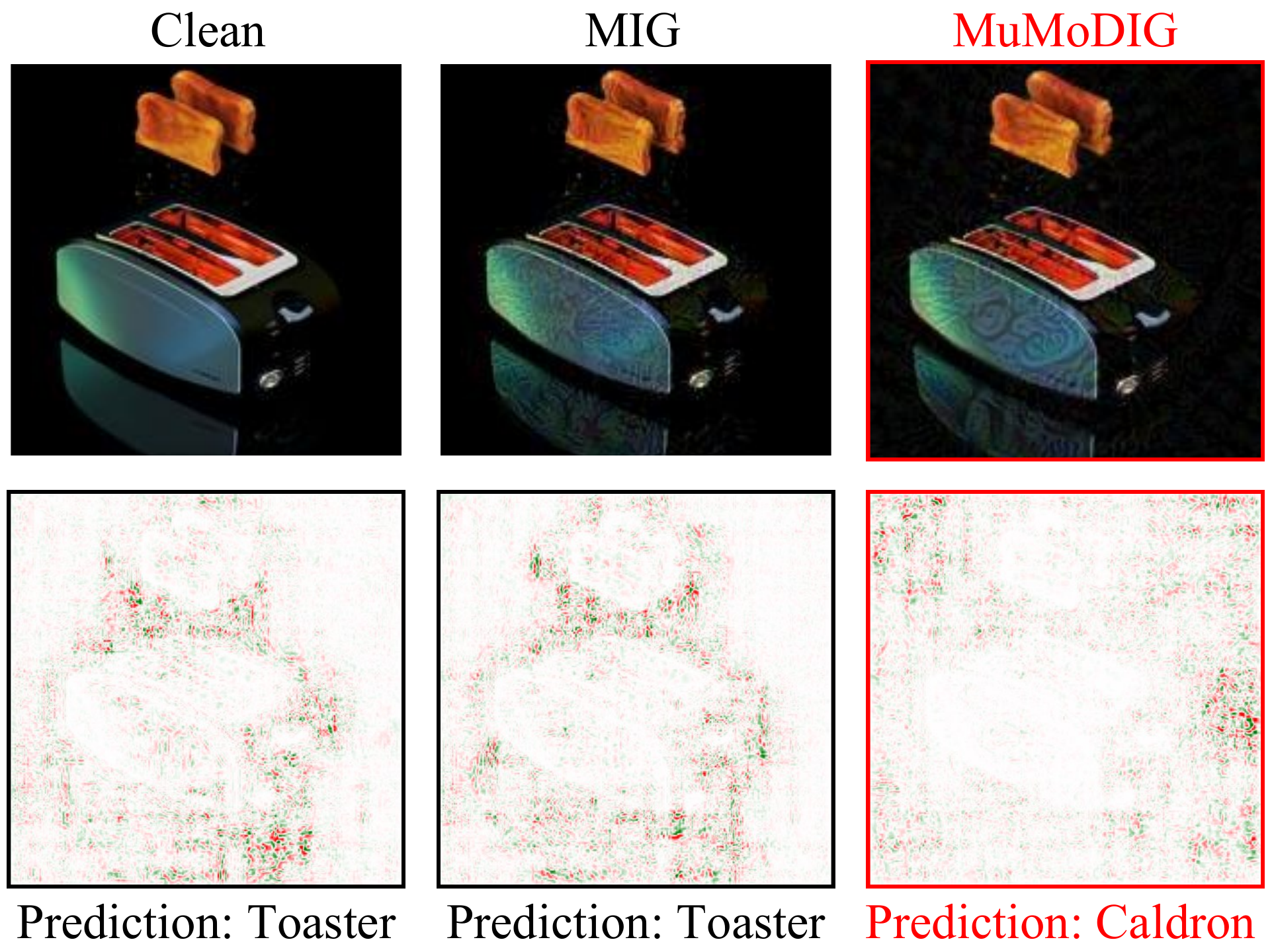}
\caption{Model attribution results for adversarial examples generated by our MuMoDIG vs. MIG \cite{mig} on the target model PiT-T.
% which attributes the target model's top-1 prediction to the input image using integrated gradients (IG) \cite{ig}. 
MuMoDIG concentrates more on the background, showcasing its superiority in disrupting the model prediction. Here, RN-18 is the surrogate model.}
\label{ig_visual}
\end{figure}

Recently, several transferable attacks \cite{taig,mig} have migrated the idea of integrated gradients (IG) \cite{ig} from the task of DNN interpretability to transferable attacks. 
Originally, IG interprets a DNN model by attributing its predictions to its input features along a given path.
In the context of transferable attacks, the accumulation effect of IG leads to a stable update direction of adversarial gradients and substantially enhances transferability.
In particular, existing IG-based attacks naively adopt IG implementations that rely on a single path.

In this paper, we demonstrate that, however, such a naive adoption has largely limited the transferability of IG-based attacks, since the calculated gradients may be easily affected by the high curvature position of the output manifold \cite{gig}. 
To address this limitation, we propose to refine the integration path in three key aspects: multiplicity, monotonicity, and diversity. 
First, we conduct theoretical analyses to show that directly adopting arbitrary multiple paths even harms the transferability due to the conflict between integration paths and gradients.
Instead, we should ensure the monotonicity of the paths, which is especially achieved by a newly proposed Lower Bound Quantization (LBQ) method.
Furthermore, based on our finding that the cosine similarity of interpolated points along the path is too high to avoid overfitting, we propose to diversify the paths based on composite random transformation.

Following the above steps, we finally propose the Multiple Monotonic Diversified Integrated Gradients (MuMoDIG) attack, which can generate highly transferable adversarial examples.
As illustrated in Figure \ref{ig_visual}, the model attribution of adversarial examples generated by our MuMoDIG focuses more on the image background than the foreground objects.
This explains the stronger disruption of our adversarial examples on the target model's predictions than the latest IG-based attack, MIG \cite{mig}.

In sum, we make the following main contributions:
\begin{itemize}

    \item We demonstrate that existing IG-based attacks suffer from limited transferability since they naively adopt IG with a single integration path.
     We address this limitation by refining the integration path in three key aspects: multiplicity, monotonicity, and diversity.

    \item We propose a new transferable attack called the Multiple Monotonic Diversified Integrated Gradients (MuMoDIG), and we conduct all-sided analyses to support its design. In particular, to enforce the monotonicity, we propose a Lower Bound Quantization (LBQ) method.

    \item We validate the superiority of MuMoDIG through extensive experiments on ImageNet against various CNN and ViT models and popular defenses.
    The results show that MuMoDIG outperforms the latest IG-based attack by up to 37.3\% and other advanced attacks by 8.4\%.

\end{itemize}

\section{Related Work}
% In this section, we review representative transferable attack methods and Integrated Gradients (IG)-based attribution methods, both of which are necessary to understand our work on IG-based transferable attacks.

\subsection{Transferable Attacks}
Transferable attacks have been extensively studied due to their practicality. 
% They are typically built upon the iterative attack Basic Iterative Method (BIM) \cite{bim}, which iteratively updates adversarial examples by following the direction of the gradient sign.
% Specifically, 
Momentum Iterative Method (MIM) \cite{mi} incorporates a momentum term in each iteration. Diversity Input Method (DIM) employs padding and resizing input transformations. Skip Gradient Method (SGM) \cite{sgm} scales the gradients passing through the residual modules, reducing the overfitting to the surrogate model. Pay No Attention and Patch Out (PNAPO) \cite{pna} truncates gradients from the attention branch and randomly masks the input patch to mitigate overfitting. 
See \citet{zhao2023revisiting} for a detailed review of five typical categories of transferable attacks.

A common practice to further improve transferability is by combining different operations, such as multiple image transformations and gradient modifications.
Specifically, Spectrum Simulation Attack (SSA) \cite{ssa} combines random Gaussian noise and spectrum masking. Gradient Relevance Attack (GRA) \cite{gra} adjusts gradient direction with the aid of noisy inputs, utilizing a gradient relevance framework and decay indicator. Structure Invariant Attack (SIA) \cite{sia} applies a library of ten different image transformations to various local regions of the input.

\subsection{Integrated Gradients}
% Gradient-based attribution methods demonstrate outstanding performance in the field of model interpretability, with visual results that closely align with the human vision system. 
Gradient-based attribution methods excel in model interpretability, producing visual results aligned with human vision.
Integrated Gradients (IG) \cite{ig} attribute model's prediction to the input along a straight line from a given black image baseline to the input. Blur Integrated Gradients (BlurIG) \cite{big} considers the path generated by a sequence of blurred image baselines. Guided Integrated Gradients (GIG) \cite{gig} posits that the high curvature of a DNN's output manifold results in larger gradients, which markedly influence the final attribution values. To address this problem, GIG employs an adaptive path method. Important Direction Gradient Integration (IDGI) \cite{idig} further improves the consistency with human visual experience, separating the original gradients into noisy gradients and important gradients, and using only the important gradients for attribution.

Recently, several studies have introduced IG to transferable attacks~\cite{taig,mig}.
% They consider interpolated points along the integration path as auxiliary examples for more stable gradient computation.
% Among them, 
The latest MIG \cite{mig} achieves high transferability by incorporating a momentum term.
Our work follows this research direction and specifically addresses the problem of existing IG-based attacks by refining their integration paths.

\begin{figure*}[!t]
\centering
    \includegraphics[width=2.0\columnwidth]{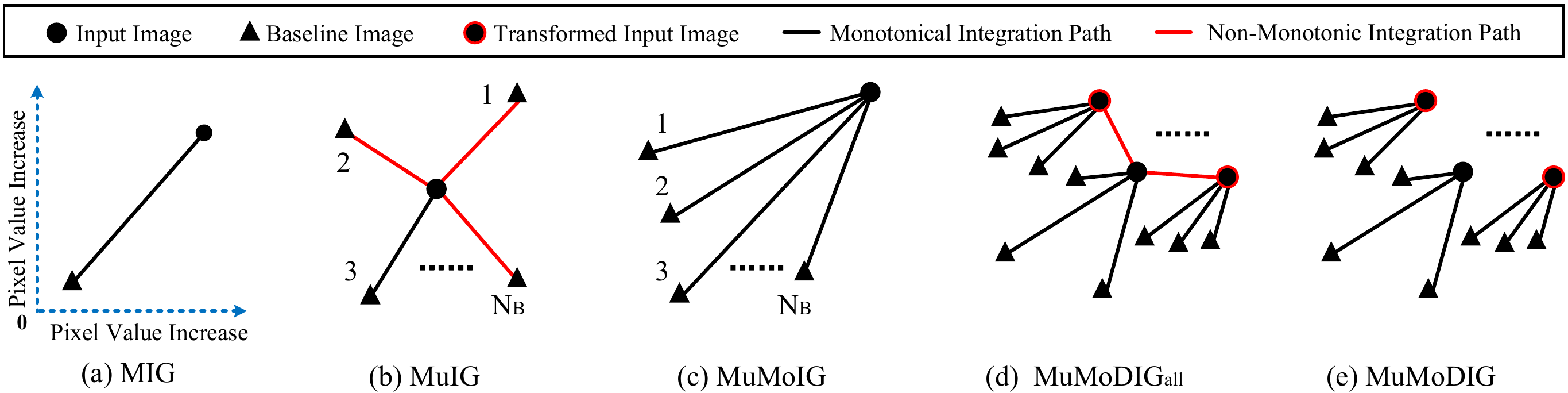}
\caption{(a) MIG \cite{mig} adopts a single integration path with a black image baseline. Our (b) MuIG adopts multiple integration paths with arbitrary baselines, with (c) MuMoIG further enforcing monotonicity, (d) MuMoDIG$_\textrm{{all}}$ diversifying paths and keeping all without enforcing their monotonicity, and (e) MuMoDIG removing non-monotonic diversified paths.}
\label{5methods}
\end{figure*}
\section{Methodology}

% This section will provide essential preliminaries and detail our methods.

\subsection{Integrated Gradients in Transferable Attacks}
Given a clean image $x$ with channel $C$, height $H$, width $W$, true label $y$, transferable attacks optimize an adversarial perturbation $\delta'$ bounded by $\epsilon$ on a surrogate model $f$. This process can be formulated as:
\begin{equation}
    \label{goal}
    \delta' = \mathop {\arg \max }\limits_\delta  L(f(x + \delta ),y), \,\\
    s.t. \,\, {\left\| \delta  \right\|_\infty } \le \epsilon,
\end{equation}
where $f(x)$ denotes the softmax output, and $L$ is the loss function, typically the cross-entropy loss. $\epsilon$ is commonly an $L_p$ norm and also other percetual measures are explored~\cite{perc,advcolor,chen2024content}.

Integrated gradients (IG) is a tensor with the same dimension as the input image.
Given a surrogate model $f$, the $i$-th element of IG during generating the adversarial perturbation can be obtained by:
\begin{equation}
    \label{ig}
IG{({x_t})_i} = ({({x_t})_i} - {b_i}) \cdot \int_0^1 {\frac{{\partial f(b + \alpha  \cdot ({x_t} - b))}}{{\partial {{({x_t})}_i}}}d\alpha },
\end{equation}
where $x_t=x+\delta_t$ is the input image at the $t$-th iteration, the baseline $b$ is typically a black image, and $(({x_t}) - {b})$ is a straight integration path.
Eq. (\ref{ig}) can be approximated by:
\begin{equation}
\label{ig_ap}
IG{({x_t})_i} \approx \frac{{({{({x_t})}_i} - {b_i})}}{{{N_I}}} \cdot \sum\limits_{k = 0}^{{N_I} - 1} {\frac{{\partial f(b + \frac{{k + \lambda }}{{{N_I}}} \cdot ({x_t} - b))}}{{\partial {{({x_t})}_i}}}},
\end{equation}
where ${{N_I}}$ is the number of interpolation points, and the position factor $\lambda \in [0,1]$ is introduced to control the position of interpolated points in each segment of the straight path.

\begin{figure}[!t]
\centering
    \includegraphics[width=0.7\columnwidth]{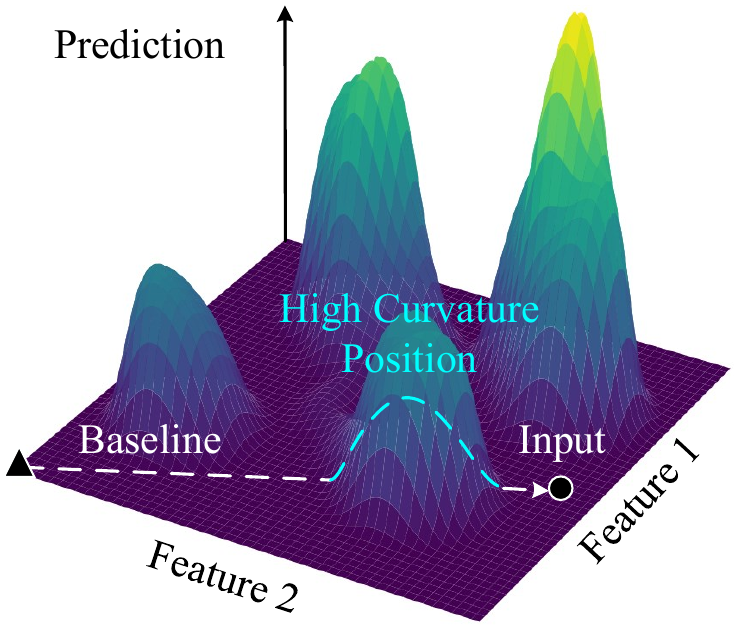}
\caption{The influence of the output manifold's high curvature position towards the single integration path.}
\label{cur_problem}
\end{figure}

% \subsection{Motivation}
In the following, we present our transferable attack method that refines the existing IG-based method by gradually achieving multiple, monotonic, and diversified integration paths, as illustrated in Figure \ref{5methods}.

\subsection{Attack with Multiple Monotonic Integration Paths}
Existing work on model interpretability demonstrates that IG with only one integration path may accumulate unstable gradients due to the high curvature position of the output manifold \cite{gig}, as illustrated in Figure \ref{cur_problem}.
We suppose this problem would also limit the direct use of IG in transferable attacks.
Therefore, we propose the Multiple Integrated Gradients (MuIG), assuming that the baseline $b$ follows a certain distribution rather than a fixed black image as before \cite{taig,mig}.
To this end, the $i$-th element of MuIG becomes:
\begin{equation}
\label{eig}
MuIG{({x_t})_i} = {{\rm E}_b}(IG{({x_t})_i}),
\end{equation}
which adds the expectation operation ${\rm E}_b$ to Eq. \ref{ig}.

When utilizing multiple integration paths to interpret DNNs, there are numerous options for baseline selection \cite{ig_summary}, such as solid color images, noisy images, and blurred images \cite{big}. However, our exploratory experiments show that using arbitrary baselines severely harms transferability as the number of baselines increases.
This is because in this case, the sign of the integration path $({x_t} - {b})$ does not necessarily align with the sign of the gradient $\frac{{\partial f({x_t} + \alpha \cdot ({x_t} - b))}}{{\partial {x_t}}}$.
As a result, when the signs of corresponding elements between the two are completely opposite, the gradient direction will be most severely disrupted, leading to opposite update direction and deteriorating transferability.

To avoid such a conflict between the integration path and gradient, we define the Monotonic Integration Path and give a proposition (see proof in Appendix) as follows:
\begin{definition}[Monotonic Integration Path]
\label{def:miip}
Consider an integration path consisting of a sequence of interpolated points $({x_0},...,{x_{{N_I}-1}})$. For any interpolated point $x_k$, if the following conditions hold: 1) $\forall s < k, \, ({x_s})_i \leq ({x_k})_i$; 2) $\forall m > k, \, ({x_m})_i \geq ({x_k})_i$, where $0 \le s < k < m \le {N_I} - 1$ and $0 \le i \le C \cdot H \cdot W$, then this path is called a Monotonic Integration Path.
\end{definition}
\begin{proposition}
\label{pro:miip}
The integration path should be a Monotonic Integration Path when using integrated gradients to generate adversarial examples in transferable attacks. 
\end{proposition}

Definition \ref{def:miip} ensures that the elements of earlier interpolated points in the sequence are always less than or equal to the corresponding elements of the later interpolated points. Proposition \ref{pro:miip} engages that the signs of the elements in the gradients during the adversarial example generation process are not altered by the integration path, since all elements of $({x_t} - b)$ are positive. 
Definition \ref{def:miip} also explains the enhanced transferability by using a black image baseline, as in the existing method, MIG \cite{mig}, which naturally forms a monotonic integration path.
In contrast, using integration paths constructed by arbitrary baselines, such as noisy images, does not adhere to the above definition.

\begin{figure}[!t]
\centering
    \includegraphics[width=1.0\columnwidth]{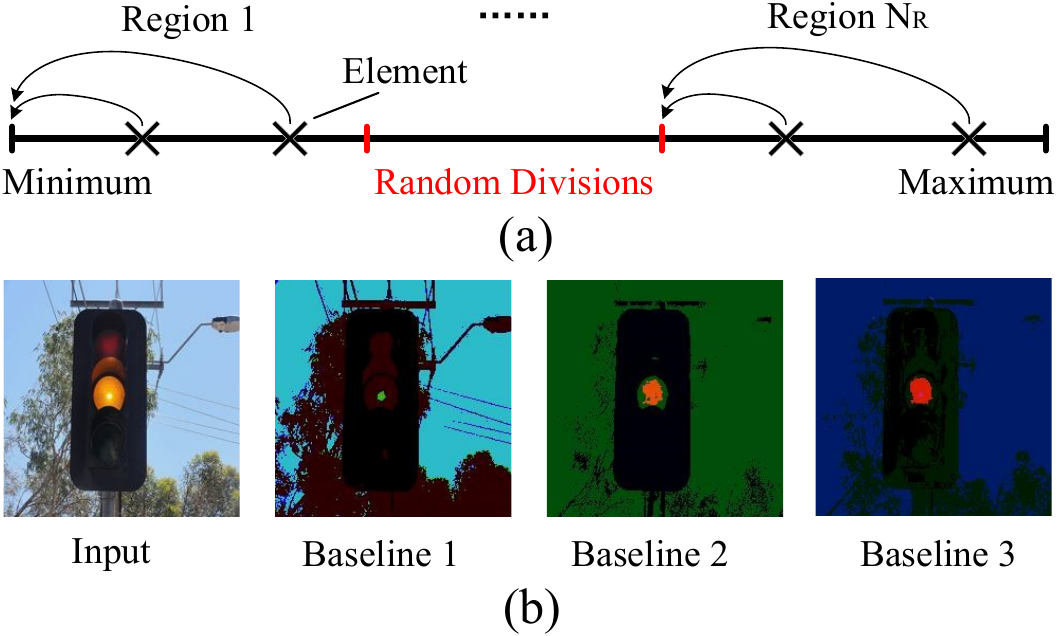}
\caption{(a) Lower Bound Quantization (LBQ) quantizes all elements in each region to their minimum value, resulting in (b) baseline images that enforce monotonic paths.}
\label{lbq}
\end{figure}

Inspired by the Randomized Quantization method \cite{rq}, we propose a Lower Bound Quantization (LBQ) method to generate baselines that conform to Proposition \ref{pro:miip}, \ie, all elements of the baseline $LBQ({x_t})$ are less than or equal to $x_t$.
Specifically, LBQ is implemented as follows:
 \begin{enumerate}
  \item Convert each channel of $ x_t $ into a one-dimensional vector, with the elements of each channel's vector sorted in ascending order of value.

  \item Randomly select $ N_R - 1 $ ($2 \le {N_R}$) divisions to split each vector into $ N_R $ regions and replace all elements in each region with the minimum value of that region.

  \item Convert the above-processed vectors of each channel back to the original dimensions of each channel.
\end{enumerate}
Figure \ref{lbq} illustrates LBQ and visualizes some examples of the generated image baselines.
We term MuIG with the above monotonic integration paths as Multiple Monotonic Integrated Gradients (MuMoIG).

\begin{table*}[!t]
\centering
\setlength{\tabcolsep}{0.9mm}
\small
\begin{tabular}{ccccccccccccccc}
\toprule[1pt]
Attack     & RN-18 & RN-101 & RNX-50 & DN-121 & ViT-B & PiT-B & Vis-S & Swin-T & AT   & HGD  & NRP  & Bit  & JPEG & Mean \\ \midrule[1pt]
MIG        & 100.0* & 55.8   & 62.3   & 87.5   & 20.6  & 33.6  & 45.8  & 49.8   & 45.8 & 46.3 & 42.5 & 60.2 & 59.0 & 50.8 \\ \hline
MuIG (${N_B}=6,{N_I}=6$)    & 18.6*  & 6.6    & 9.3    & 14.1   & 3.5   & 5.7   & 6.8   & 9.1    & 40.7 & 5.8  & 23.9 & 13.3 & 23.8 & 13.6 \\
MuMoIG (${N_B}=1,{N_I}=6$)    & 100.0* & 59.2   & 64.3   & 91.4   & 23.1  & 35.3  & 49.3  & 51.7   & 45.4 & 50.0 & 42.7 & 60.8 & 62.6 & 53.0 \\
MuMoIG (${N_B}=6,{N_I}=6$)    & 100.0* & 64.4   & 70.5   & 94.3   & 25.9  & 36.2  & 53.4  & 55.4   & 46.6 & 54.4 & 45.0 & 64.5 & 66.1 & 56.4 \\
MuMoDIG$_\textrm{{all}}$ & 99.8*  & 70.6   & 74.0   & 94.6   & 33.6  & 45.5  & 61.6  & 61.0   & 47.6 & 64.4 & 47.8 & 68.4 & 69.9 & 61.6 \\
MuMoDIG &
  \textbf{100.0}* &
  \textbf{85.3} &
  \textbf{86.9} &
  \textbf{97.8} &
  \textbf{43.5} &
  \textbf{57.0} &
  \textbf{73.6} &
  \textbf{75.1} &
  \textbf{51.9} &
  \textbf{89.7} &
  \textbf{60.7} &
  \textbf{85.2} &
  \textbf{80.8} &
  \textbf{73.9} \\ \bottomrule[1pt]
\end{tabular}
\caption{Attack success rates ($\%$) of gradually refining MIG \cite{mig} to form our ultimate attack, MuMoDIG.
The surrogate model is RN-18, and the ``Mean'' column excludes the white-box results marked with *.}
\label{table1}
\end{table*}

\subsection{Attack with Diversified Integration Paths}

\begin{figure}[!t]
\centering
    \includegraphics[width=0.75\columnwidth]{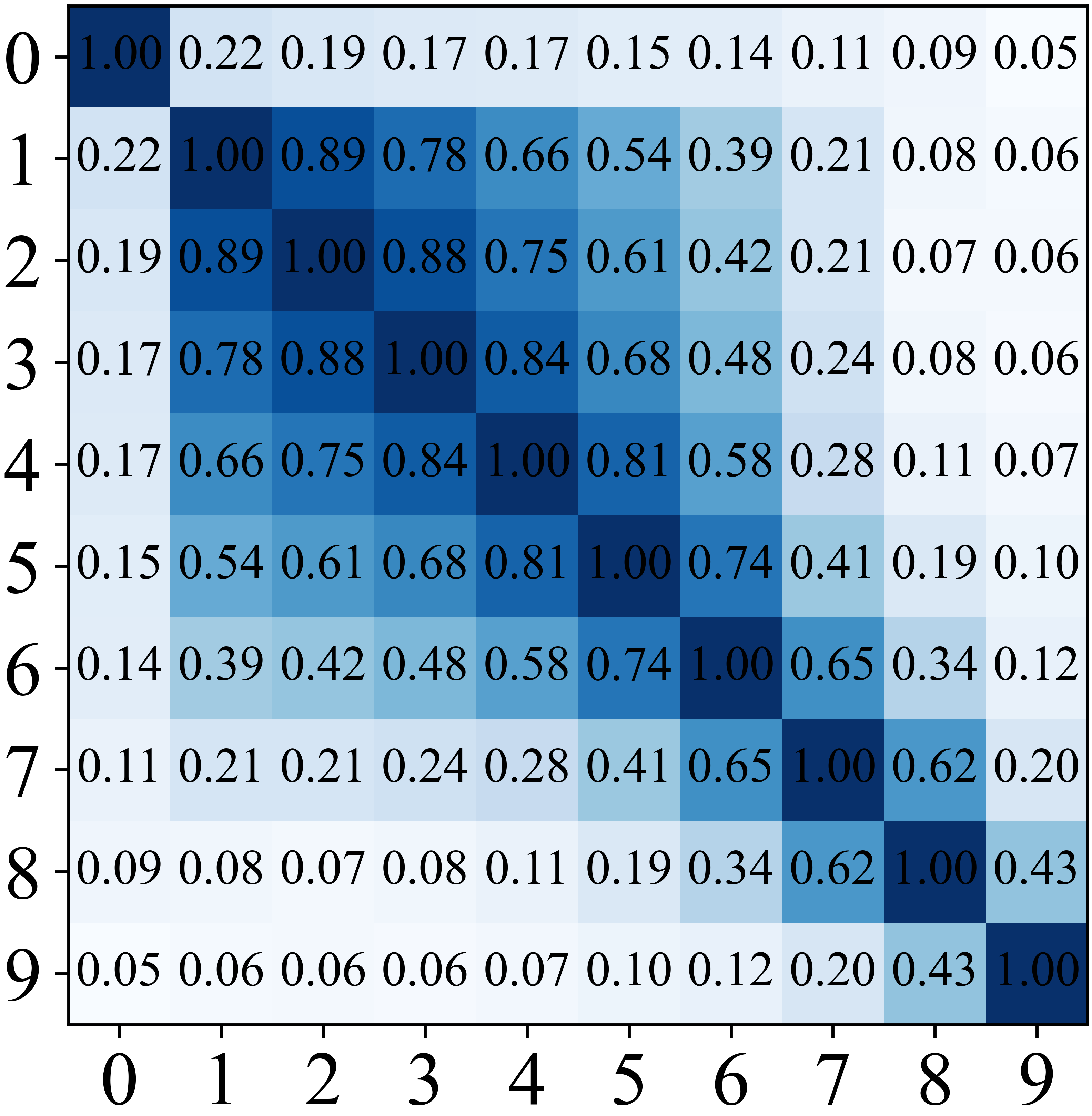}
\caption{The cosine similarity calculated between the gradients at 10 interpolated points along a straight path.}
\label{diver_problem}
\end{figure}

After ensuring the monotonicity of multiple integration paths, we further look at the properties of interpolated points along the path.
Specifically, we compute the cosine similarity of interpolated points. Figure \ref{diver_problem} shows that gradients calculated at a sequence of interpolated points lack diversity, as the cosine similarity between the gradients at adjacent positions is very high.
Accumulating such similar gradients using Eq. \ref{ig_ap} cannot effectively reduce the overfitting to the surrogate model $f$, because these gradients provide limited information in the high-dimensional space.

To address this limitation, we follow previous work \cite{little} to apply input transformations to reduce the cosine similarity of gradients as follows:
\begin{equation}
\label{deig}
MuMoDI{G_\textrm{{all}}}{({x_t})_i} = {{\rm{E}}_T}({{\rm{E}}_{b|T}}(I{G_\textrm{{all}}}{({x_t})_i})),
\end{equation}
where $T$ denotes the input transformations, and all diversified paths are kept no matter if they satisfy monotonicity. This forms the Multiple Monotonic Diversified Integrated Gradients$_\textrm{{all}}$ (MuMoDIG$_\textrm{{all}}$).
To ensure sufficient gradient diversity, the transformation typically contains composite operations, from which one operation is randomly selected each time.
Here, we follow the common practice of using two simple transformations: the resizing and padding (RP)~\cite{di} and the affine transformation (AF), selected with equal probability.

%\toprule[1pt]
%\midrule[1pt]
%\bottomrule[1pt]

\begin{table*}[!t]
\centering
\setlength{\tabcolsep}{1.2mm}
\small
\begin{tabular}{cccccccccccccccc}
\toprule[1pt]
Surrogate & Attack & RN-18 & RN-101 & RNX-50 & DN-121 & ViT-B & PiT-B & Vis-S & Swin-T & AT & HGD & NRP & Bit &JPEG & Mean \\ \midrule[1pt]
\multirow{5}{*}{RN-18} & MIG & \textbf{100.0*} & 55.8 & 62.3 & 87.5 & 20.6 & 33.6 & 45.8 & 49.8 & 45.8 & 46.3 & 42.5 & 60.2 & 59.0 & 50.8 \\
 & GRA & \textbf{100.0*} & 58.2 & 64.0 & 90.9 & 25.3 & 33.3 & 47.2 & 56.9 & 50.0 & 51.5 & 49.6 & 64.0 & 64.1 & 54.6 \\
 & SSA & \textbf{100.0*} & 58.3 & 62.8 & 90.9 & 25.0 & 35.9 & 47.2 & 53.0 & 47.1 & 50.8 & 48.3 & 63.7 & 66.7 & 54.1 \\
 & SIA & \textbf{100.0*} & 81.2 & 84.6 & \textbf{98.2} & 36.9 & 50.4 & 70.1 & 71.4 & 47.6 & 70.0 & 49.8 & 75.5 & 70.4 & \underline{67.2} \\
 & MuMoDIG & \textbf{100.0*} & \textbf{85.3} & \textbf{86.9} & 97.8 & \textbf{43.5} & \textbf{57.0} & \textbf{73.6} & \textbf{75.1} & \textbf{51.9} & \textbf{89.7} & \textbf{60.7} & \textbf{85.2} & \textbf{80.8} & \textbf{73.9} \\ \hline
\multirow{5}{*}{DN-121} & MIG & 85.1 & 65.6 & 69.7 & \textbf{100.0*} & 30.1 & 40.9 & 55.9 & 54.9 & 44.6 & 58.2 & 45.3 & 64.2 & 61.1 & 57.5 \\
 & GRA & 95.3 & 82.5 & 85.0 & \textbf{100.0*} & 46.9 & 58.5 & 70.7 & 71.5 & \textbf{53.5} & 80.6 & \textbf{64.6} & 78.6 & 77.9 & \underline{72.5} \\
 & SSA & 94.8 & 83.0 & 83.8 & \textbf{100.0*} & 43.5 & 56.1 & 71.1 & 70.4 & 47.6 & 77.8 & 60.4 & 77.1 & 77.3 & 70.7 \\
 & SIA & \textbf{97.2} & 86.7 & 90.1 & \textbf{100.0*} & 43.5 & 59.6 & 77.4 & 73.4 & 47.5 & 80.9 & 53.8 & 79.6 & 72.8 & 72.1 \\
 & MuMoDIG & \textbf{97.2} & \textbf{88.7} & \textbf{90.2} & \textbf{100.0*} & \textbf{49.1} & \textbf{63.3} & \textbf{78.8} & \textbf{75.5} & 47.8 & \textbf{87.9} & 57.8 & \textbf{82.4} & \textbf{78.2} & \textbf{75.0} \\ \hline
\multirow{5}{*}{MN-v3} & MIG & 69.2 & 30.0 & 35.8 & 60.0 & 20.1 & 26.7 & 35.0 & 39.8 & 46.8 & 46.3 & 34.3 & 45.0 & 48.7 & 41.4 \\
 & GRA & 75.0 & 32.6 & 38.7 & 61.5 & 21.4 & 27.8 & 35.0 & 43.2 & 50.8 & 51.5 & 38.9 & 49.3 & 54.5 & 44.6 \\
 & SSA & 71.6 & 28.1 & 35.9 & 61.4 & 17.6 & 24.4 & 31.7 & 38.4 & 49.1 & 50.8 & 37.2 & 46.3 & 53.9 & 42.0 \\
 & SIA & 84.4 & 43.4 & 48.8 & 75.9 & 26.9 & 37.2 & 48.4 & 52.2 & 50.3 & 70.0 & 40.6 & 57.9 & 56.9 & \underline{53.3} \\
 & MuMoDIG & \textbf{88.6} & \textbf{52.4} & \textbf{57.0} & \textbf{80.9} & \textbf{36.4} & \textbf{46.8} & \textbf{57.2} & \textbf{62.4} & \textbf{51.7} & \textbf{89.7} & \textbf{47.3} & \textbf{66.2} & \textbf{65.4} & \textbf{61.7} \\ \hline
\multirow{5}{*}{PiT-T} & MIG & 70.0 & 38.6 & 44.3 & 64.5 & 32.2 & 43.5 & 46.0 & 56.6 & 45.9 & 35.7 & 37.8 & 54.3 & 55.4 & 48.1 \\
 & GRA & 86.9 & 61.5 & 66.2 & 82.2 & 57.1 & 68.2 & 68.9 & 77.8 & 52.6 & 58.2 & 57.9 & 74.3 & 72.0 & 68.0 \\
 & SSA & 84.2 & 55.7 & 61.2 & 77.8 & 46.3 & 60.2 & 62.4 & 72.0 & 49.4 & 53.6 & 51.3 & 68.9 & 71.1 & 62.6 \\
 & SIA & \textbf{93.7} & 71.9 & 76.7 & 88.8 & 64.2 & 77.2 & 79.8 & 84.3 & 51.8 & 64.2 & 55.0 & 80.7 & 74.6 & \underline{74.1} \\
 & MuMoDIG & 92.5 & \textbf{74.8} & \textbf{79.0} & \textbf{90.5} & \textbf{69.5} & \textbf{80.3} & \textbf{80.9} & \textbf{86.9} & \textbf{53.5} & \textbf{73.9} & \textbf{59.3} & \textbf{83.4} & \textbf{78.4} & \textbf{77.1} \\ \hline
\multirow{5}{*}{DeiT-T} & MIG & 66.4 & 36.0 & 41.6 & 61.6 & 56.2 & 39.6 & 46.3 & 68.6 & 45.3 & 31.3 & 39.8 & 56.3 & 57.0 & 49.7 \\
 & GRA & 81.2 & 52.6 & 59.5 & 77.1 & 77.0 & 60.5 & 65.8 & 83.6 & 50.9 & 49.7 & 56.2 & 73.2 & 72.5 & 66.1 \\
 & SSA & 79.0 & 48.2 & 54.2 & 73.0 & 70.9 & 52.6 & 58.9 & 78.5 & 49.3 & 45.6 & 50.2 & 68.5 & 71.3 & 61.6 \\
 & SIA & 90.4 & 70.0 & 75.8 & 87.8 & 83.3 & 79.8 & 83.5 & 90.2 & 52.0 & 64.8 & 58.3 & 82.8 & 77.2 & \underline{76.6} \\
 & MuMoDIG & \textbf{91.6} & \textbf{73.8} & \textbf{77.5} & \textbf{89.3} & \textbf{83.8} & \textbf{82.4} & \textbf{84.8} & \textbf{90.9} & \textbf{53.8} & \textbf{73.4} & \textbf{63.2} & \textbf{84.2} & \textbf{80.8} & \textbf{79.2} \\ \hline
\multirow{5}{*}{Swin-T} & MIG & 47.2 & 25.5 & 32.4 & 42.1 & 23.7 & 33.5 & 41.9 & 98.8* & 42.9 & 19.9 & 28.7 & 45.7 & 45.2 & 35.7 \\
 & GRA & \textbf{83.0} & 69.9 & 74.4 & 80.8 & \textbf{73.7} & 81.3 & 84.4 & 98.6* & \textbf{51.8} & 66.1 & \textbf{69.1} & 80.5 & \textbf{78.4} & \textbf{74.5} \\
 & SSA & 80.4 & 66.2 & 70.2 & 79.8 & 64.3 & 75.1 & 82.4 & \textbf{98.9*} & 47.5 & 61.9 & 62.5 & 78.2 & 77.7 & 70.5 \\
 & SIA & 79.3 & 67.1 & 72.7 & 80.8 & 59.0 & 79.5 & 85.2 & 98.7* & 45.3 & 54.9 & 48.2 & 77.0 & 68.1 & 68.1 \\
 & MuMoDIG & 82.8 & \textbf{73.4} & \textbf{76.9} & \textbf{84.3} & 67.6 & \textbf{82.7} & \textbf{85.2} & 98.3* & 46.9 & \textbf{66.3} & 53.5 & \textbf{82.0} & 73.9 & \underline{73.0} \\ \bottomrule[1pt]
\end{tabular}
\caption{Attack success rates ($\%$) of our MuMoDIG vs. state-of-the-art transformation-based attacks. For defenses, AT uses RN-50, HGD uses its default setting, and NRP, JPEG, and Bit results are averaged on eight target models. The results with underlined in the "Mean" columns are the second best, and the ``Mean'' column excludes the white-box results marked with *.}
\label{table2}
\end{table*}

Furthermore, it is easy to notice that the integration paths from transformed inputs to the input cannot be guaranteed to be monotonic, violating our Proposition \ref{pro:miip}.
Therefore, we discard such non-monotonic paths, and Eq. \ref{deig} becomes:
\begin{equation}
\label{MuMoDIG}
MuMoDIG{({x_t})_i} = {{\rm E}_T}({{\rm E}_{b|T}}(IG_\textrm{{mo}}{({x_t})_i})),
\end{equation}
which can be approximated with Monte Carlo Sampling by:
\begin{equation}
\label{depig_ap}
MuMoDIG{({x_t})_i} \approx I \cdot \sum\limits_{p = 0}^{{N_T}} {\sum\limits_{q = 0}^{{N_B} - 1} {{{(IG_\textrm{{mo}}{{({x_t})}_{p,q}})}_i}} } ,
\end{equation}
where $I = \frac{1}{{({N_T} + 1) \cdot {N_B}}}$, with $N_T$ as the number of sampled transformations and $N_B$ as the number of sampled baselines.
Here, a fixed identity transformation is also applied to the input image, except for the $N_T$ extra transformations. 
After all the above steps, we achieve the ultimate version of our method, Multiple Monotonic Diversified Integrated Gradients (MuMoDIG).

Similar to MIG \cite{mig}, our MuMoDIG is also integrated with the momentum term \cite{mi} to further boost transferability.
In this case, the above $I$ can be omitted as it cancels out during gradient normalization.
We also find that using $ -\log f(x)$ instead of $-f(x)$ for the loss function yields a slightly better performance for both our and existing IG-based attacks.
%\begin{equation}
%\label{mi}
%\left\{ \begin{array}{l}
%{g_{t + 1}} = \mu  \cdot {g_t} + \frac{{MuMoDIG({x_t})}}{{{{\left\| %{MuMoDIG({x_t})} \right\|}_1}}}\\
%{x_{t + 1}} = Cli{p_\varepsilon }\{ {x_t} + \alpha  \cdot sign({g_{t + 1}})\}
%\end{array} \right.,
%\end{equation}
% where $\mu$ is the decay factor, $\alpha$ is the step size, ${g_t}$ is the momentum term at the $t$-th iteration and $Cli{p_\varepsilon}$ limits the change of each pixel within $\varepsilon$.

\subsection{IG for Interpretability vs. for Transferabability}

% Given the above descriptions, we have realized that making IG suitable for transferable attacks is a non-trivial task.
Here, we shed more light on the fundamental differences between using IG in model interpretability and transferable attacks.
In general, IG in model interpretability focuses on providing better visual explanations by prioritizing the overall magnitude of the product between gradients and the integration path.
Differently, transferable attacks aim to disrupt predictions, making the sign of the gradient direction more crucial than the product's magnitude.

Specific weaknesses of directly using advanced IG techniques, such as BlurIG, GIG, and IDGI, in transferable attacks include:
1) BlurIG employs a sequence of blurred images as baselines, encountering a similar issue as discussed in MuIG: they fail to form a monotonic integration path; 2) GIG and IDGI, although with black images as baselines, face the same issue as MuMoDIG$_\textrm{{all}}$: while the straight line connecting their starting and ending points forms a monotonic integration path, the actual integration paths in between cannot guarantee monotonicity.

% \begin{itemize}
%     \item BlurIG employs a sequence of blurred images as baselines, encountering a similar issue as discussed in MuIG: they fail to form a monotonic integration path.

%     \item GIG and IDGI, although with black images as baselines, face the same issue as MuMoDIG$_\textrm{{all}}$: while the straight line connecting their starting and ending points forms a monotonic integration path, the actual integration paths in between cannot guarantee monotonicity.
% \end{itemize}

\section{Experiments}

\subsection{Experimental Settings}
\noindent\textbf{Dataset and attack baselines.}
Following many previous works \cite{vt,gra,ssa}, 
1k images from the ILSVRC2012 \cite{image} validation set are adopted in our experiments.
We compare our MuMoDIG to the latest IG-based attack, MIG \cite{mig}, as well as state-of-the-art transferable attacks with composite operations, such as SSA \cite{ssa}, GRA \cite{gra}, and SIA \cite{sia}.
All of them are equipped with multiple input transformations or special gradient modifications based on at least one transformation.
Note that Path-Augmented Method (PAM) \cite{pam} and Neuron Attribution-based Attack (NAA) \cite{naa} are path-related but not IG-based attacks, and detailed results in the Appendix show the superiority of our MuMoDIG over them.
 
\noindent\textbf{Parameters.}
Following the common practice, for all attacks, we set the maximum attack iterations as $K=10$, the maximum perturbation bound $\epsilon=16$, the step size $\alpha=1.6$, the decay factor $\mu=1.0$ in the momentum.
We set the position factor $\lambda=0.65$ and the region number ${N_R}=2$ in LBQ.
For a fair comparison, we set the number of total auxiliary inputs $N=6$ at each iteration for all attacks.
Specifically, for our MuMoDIG, we set ${N_T}=6$, ${N_B} = 1$, and ${N_I} = 1$ such that $N = {N_T} \cdot {N_B} \cdot {N_I}=6$.
All experiments are conducted on an RTX 4060 GPU with 8GB of VRAM.
Generating one image (RN-18) for MIG, GRA, SSA, SIA, and our MuMoDIG costs: 0.21s, 0.24s, 0.19s, 0.20s, and 0.28s, respectively.

%and later show that MuMoDIG is not sensitive or linearly correlated with these two parameters (see Fig.~\ref{pfrn}).

\noindent\textbf{Models and defenses.}
% Surrogate models contain three CNNs, \ie, ResNet-18 (RN-18) \cite{resnet}, DenseNet-121 (DN-121) \cite{dense} and MobileNet-v3 (MN-v3) \cite{mob}, and three ViTs, \ie, PiT-Tiny (PiT-T) \cite{pit}, Deit-Tiny (Deit-T) \cite{deit} and Swin-Tiny (Swin-T) \cite{swin}.
% Target models contain four CNNs, \ie, RN-18, RN-101, ResNeXt-50 (RNX-50) \cite{resnext}, DN-121, and four ViTs, \ie, ViT-Base (ViT-B) \cite{vit}, PiT-B, Visformer-Small (Vis-S) \cite{visf}, and Swin-T.
Surrogate models contain three CNNs, \ie, RN-18 \cite{resnet}, DN-121 \cite{dense} and MN-v3 \cite{mob}, and three ViTs, \ie, PiT-T \cite{pit}, Deit-T \cite{deit} and Swin-T \cite{swin}.
Target models contain four CNNs, \ie, RN-18, RN-101, RNX-50 \cite{resnext}, DN-121, and four ViTs, \ie, ViT-B \cite{vit}, PiT-B, Vis-S \cite{visf}, and Swin-T.

Defenses include adversarial training (AT) \cite{at},  high-level representation guided denoiser (HGD) \cite{hgd}, neural representation purifier (NRP) \cite{nrp}, Bit depth reduction (BDR) \cite{bit}, and JPEG compression \cite{jpeg}.
We also use MuMoDIG to attack the online Baidu Cloud API.

\begin{figure}[!t]
\centering
    \includegraphics[width=0.9\columnwidth]{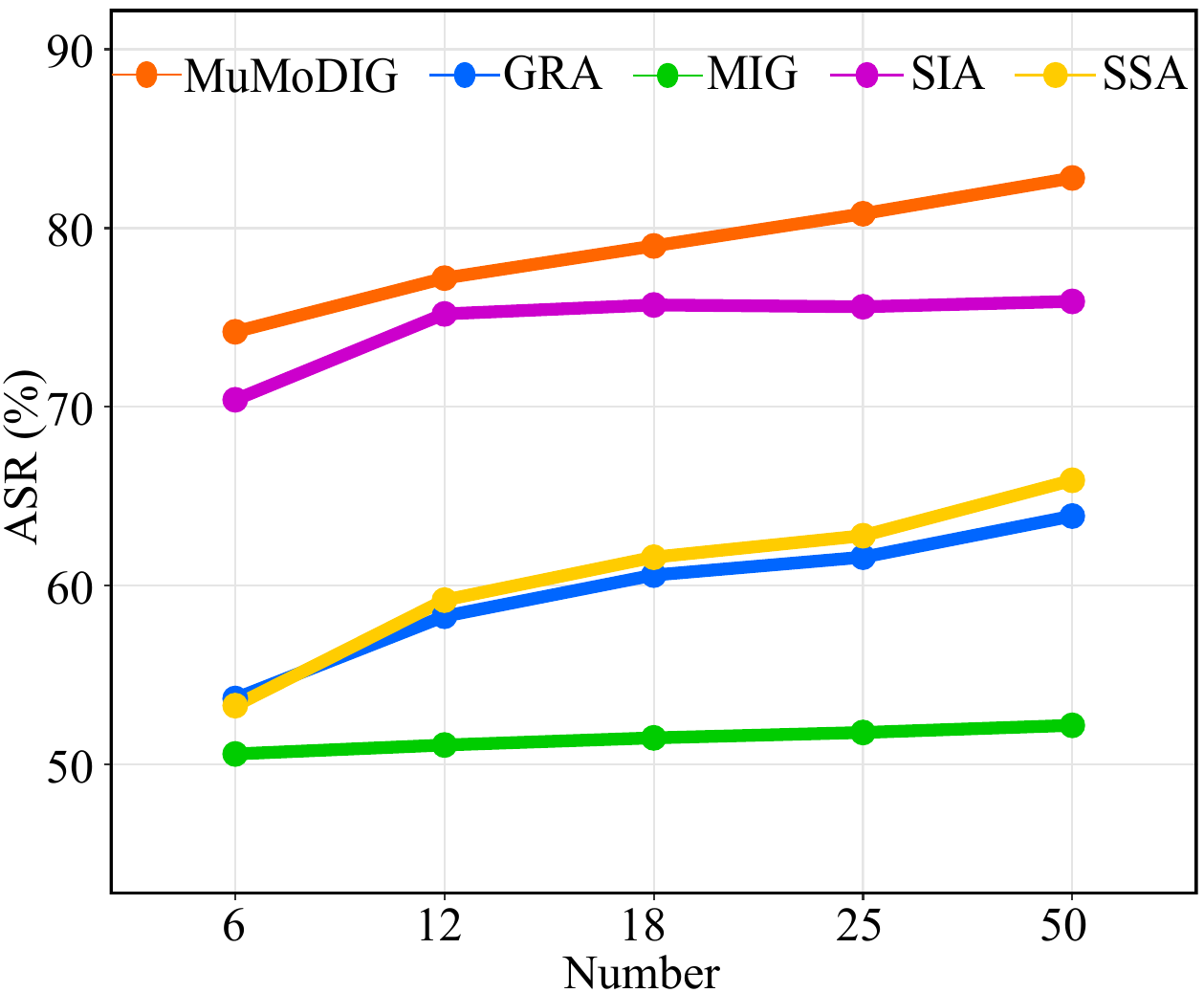}
\caption{Average success rates (\%) when adjusting the number of total auxiliary inputs $N$. The surrogate model is RN-18, and target models are the remaining seven models.}
\label{varying}
\end{figure}

\begin{table}[!t]
\centering
\begin{tabular}{cccc}
\toprule[1pt]
Surrogate & Attack&CNNs & ViTs \\ \midrule[1pt]
\multirow{3}{*}{RN-18} & SGM         & 61.0          & 36.4          \\
                       & MuMoDIG       & 90.0          & 62.3          \\
                       & MuMoDIG+SGM   & \textbf{91.1} & \textbf{66.3} \\ \hline
\multirow{3}{*}{PiT-T} & PNAPO       & 58.7          & 55.5          \\
                       & MuMoDIG       & 81.4          & 79.4          \\
                       & MuMoDIG+PNAPO & \textbf{83.4} & \textbf{82.0} \\ \bottomrule[1pt]
\end{tabular}
\caption{Attack success rates ($\%$) of integrating MuMoDIG with other attacks averaged on CNNs and ViTs.}
\label{table3}
\end{table}
\subsection{Attack Results}

\noindent\textbf{MuMoDIG vs. other IG-based attacks.} We first compare our MuMoDIG and its intermediate versions with the latest IG-based attack, MIG.
Table \ref{table1} shows the results, which fully support our claims in the above section. 
Specifically, MuIG performs the worst, indicating that using arbitrary multiple baselines even harms the transferability.
MuMoIG (${N_B}=1,{N_I}=6$) and MuMoIG (${N_B}=6,{N_I}=6$) exhibit stronger performance than MIG, highlighting the effectiveness of multiple integration paths. Lastly, MuMoDIG outperforms MuMoDIG$_\textrm{{all}}$, confirming that removing the non-monotonic paths from the diversified paths is reasonable. 

%In the following, we adopt the ultimate version MuMoDIG$_A$ as MuMoDIG.

%(a) The influence of multiple baselines crafted by LBQ, where MIG6 refers to using MIG \cite{mig} to craft adversarial examples with six interpolated points per iteration, MuIG1-6/MuIG6-6 refers to using MuIG (combined with MIM \cite{mi}) to craft adversarial examples with one/six interpolated points for each of six baselines; (b) The influence of paths from the transformed inputs to the input, where MuTIG includes complete integration paths, while MuMoDIG omits paths from transformed inputs to the input. Note that the surrogate model is RN-18 and 1,000 images are included for validation

\noindent\textbf{MuMoDIG vs. state-of-the-art transformation-based attacks.} Table \ref{table2} shows our MuMoDIG consistently outperforms other advanced attacks in almost all cases.
For example, when generating adversarial examples on MN-v3 and attacking Swin-T, our MuMoDIG achieves an attack success rate of 62.4\%, which is 10.2\% higher than the best of the other attacks.
In addition, when attacking models with defenses, MuMoDIG still outperforms others in most cases.
Figure~\ref{varying} further shows that the superiority of MuMoDIG is consistent across varied ${N}$.
Note that SIA \cite{sia} applies different transformations to specific image blocks, while our composite random transformation applies fewer transformations to the entire image.

\begin{figure}[!t]
\centering
    \includegraphics[width=1.0\columnwidth]{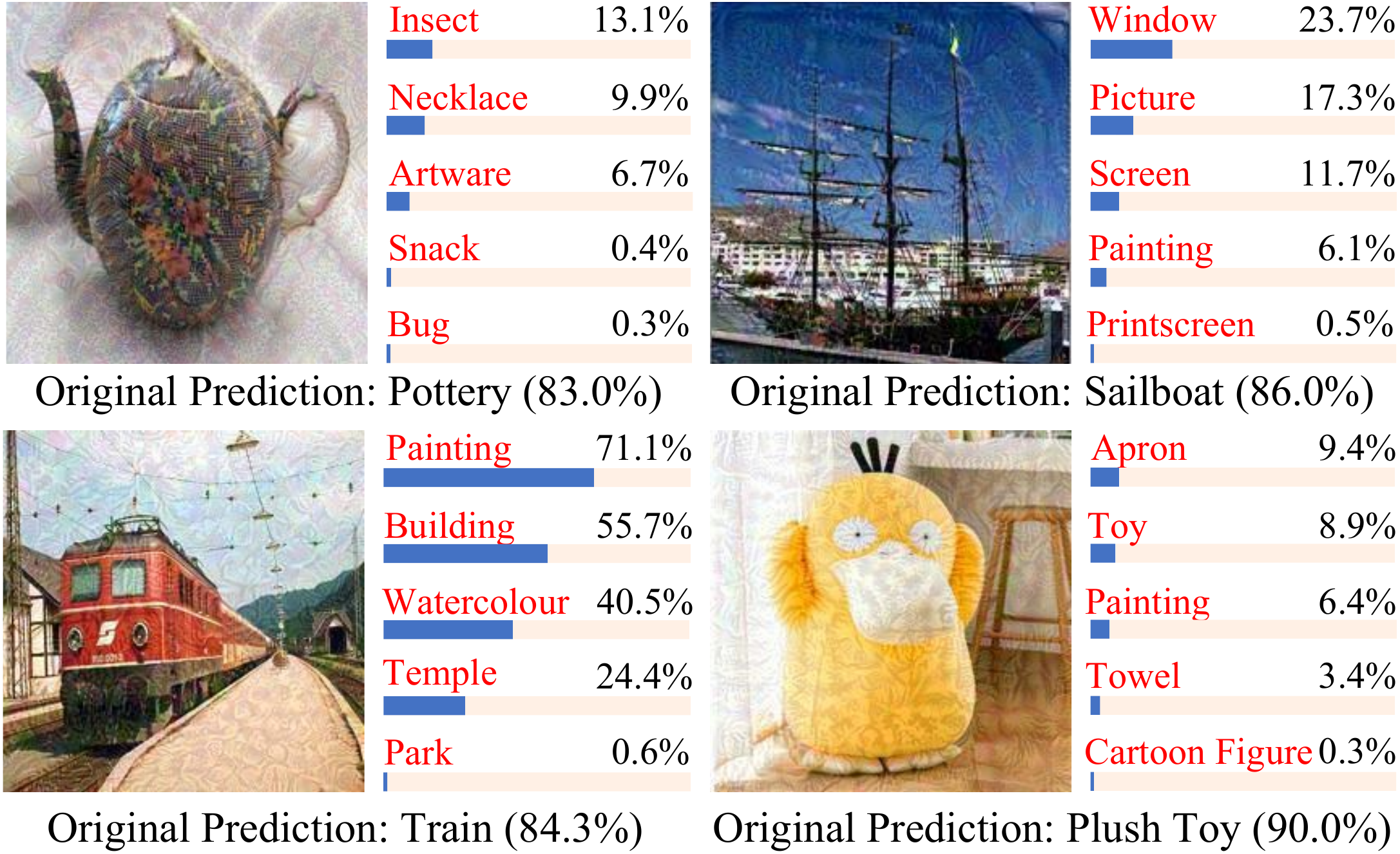} %\textwidth
\caption{Attack examples of our MuMoDIG on Baidu Cloud API. RN-18 is the surrogate model. Results on 100 images show that our MuMoDIG achieves a high ASR of 91.0\%, compared to 74.0\% for GRA and 87.0\% for SIA.}
\label{cloud}
\end{figure}

\noindent\textbf{MuMoDIG integrated with other attacks.} 
In addition to the above transformation-based attacks, we integrate our MuMoDIG with another typical type of transferable attack that refines the surrogate model.
Two such attacks are involved, with SGM \cite{sgm} targeting CNNs with residual blocks and PNAPO \cite{pna} targeting ViTs.
From Table \ref{table3}, our MuMoDIG can further boost the transferability when being integrated with these two attacks.

\noindent\textbf{MuMoDIG on attacking a real-world system.} 
%In all the above common transfer settings, there is an underlying assumption that all surrogate and target models share the same training dataset.
To test the practical usefulness of MuMoDIG, we finally consider a challenging setting with the real-world visual system, the Baidu Cloud API. 
As illustrated in Figure \ref{cloud}, for all tested examples, adding the perturbations causes the target system to make incorrect predictions, despite the original predictions being correct with high confidence.
% We also include quantitative comparisons with GRA and SIA on 100 correctly classified images, which can be found in the Appendix.

% \section{Attack on Baidu Cloud API}

% We have conducted additional quantitative attack comparisons using 100 images. The results further confirm MuMoDIG's superiority. Specifically, we selected the first 100 images correctly classified by the dataset, using RN-18 as the surrogate.

% \begin{table}[ht]
% \centering
% \caption{Performance Comparison on Baidu Cloud API}
% \begin{tabular}{lc}
% \toprule
% \textbf{Attack} & \textbf{Baidu Cloud API} \\
% \midrule
% GRA      & 74.0 \\
% SIA      & 87.0 \\
% MuMoDIG  & 91.0 \\
% \bottomrule
% \end{tabular}
% \label{tab:baidu_api_performance}
% \end{table}

\subsection{Ablation Studies}
We analyze the influence of four important parameters in MuMoDIG on the attack performance.
The surrogate is RN-18, and the target CNNs are RN-101, DN-121, and RNX-50, and ViTs are ViT-B, PiT-B, Visformer-S, and Swin-T.

\textbf{The number of total auxiliary inputs} $N$ is calculated by multiplying $N_I$, $N_B$, and $N_T$. Table \ref{table4} shows that increasing the number of either of these three types of auxiliary inputs can enhance the transferability.
% $N_B$ and $N_T$ are more effective, verifying the necessity of introducing multiple integration paths.
Specifically, increasing $N_T$ is the most effective, while increasing $N_I$ is the least effective.
This is because the pixel variations of interpolated points are less diverse, as they only reflect pixel scaling, which
is more susceptible to overfitting to the surrogate model.
Results in Table \ref{table5} further support it with different combinations of auxiliary inputs.
% Results in Table \ref{table5} further support this conclusion by showing different combinations of these three types of auxiliary inputs.
%when the number of total auxiliary inputs $N=6$.

\begin{table}[!t]
\centering
\setlength{\tabcolsep}{1mm}
% \small
\begin{tabular}{ccccccc}
\toprule[1pt]
% \multirow{2}{*}{Type} & \multicolumn{6}{c}{Number}              \\
Type/Number &1    & 6    & 12   & 18   & 25   & 50   \\ \midrule[1pt]
$N_I$                    & \textbf{59.2} & 66.8 & 67.0 & 67.1 & 66.9 & 67.9 \\
$N_B$                    & \textbf{59.2} & 69.8 & 71.7 & 72.6 & 73.1 & 75.0 \\
$N_T$                    & \textbf{59.2} & \textbf{74.2} & \textbf{77.2} & \textbf{79.0} & \textbf{80.8} & \textbf{82.8} \\ \bottomrule[1pt]
\end{tabular}
\caption{Ablation studies about the interpolation point number $N_I$, baseline number $N_B$, or sampled transformation number $N_T$. 
Results are averaged on seven target models.
When one number is varied, the others are fixed as 1.}
\label{table4}
\end{table}

\begin{table}[!t]
\centering
\begin{tabular}{ccc}
\toprule[1pt]
Type                    & CNNs          & ViTs          \\ \midrule[1pt]
$N_T=1$, $N_B=1$, $N_I=6$ & 84.7          & 53.3          \\
$N_T=1$, $N_B=2$, $N_I=3$ & 85.8          & 56.5          \\
$N_T=1$, $N_B=3$, $N_I=2$ & 87.3          & 57.0          \\
$N_T=1$, $N_B=6$, $N_I=1$ & 86.6          & 57.2          \\
$N_T=2$, $N_B=1$, $N_I=3$ & 88.1          & 57.8          \\
$N_T=2$, $N_B=3$, $N_I=1$ & 88.4          & 58.8          \\
$N_T=3$, $N_B=1$, $N_I=2$ & 89.0          & 59.9          \\
$N_T=3$, $N_B=2$, $N_I=1$ & 89.4          & 61.1          \\
$N_T=6$, $N_B=1$, $N_I=1$ & \textbf{90.0} & \textbf{62.3} \\ \bottomrule[1pt]
\end{tabular}
\caption{Ablation studies about combinations of the sampled transformation number $N_T$, baseline number $N_B$, and interpolation point number $N_I$, with $N = {N_T} \cdot {N_B} \cdot {N_I}=6$. }
\label{table5}
\end{table}

\begin{figure}[!t]
\centering
\includegraphics[width=1.0\columnwidth]{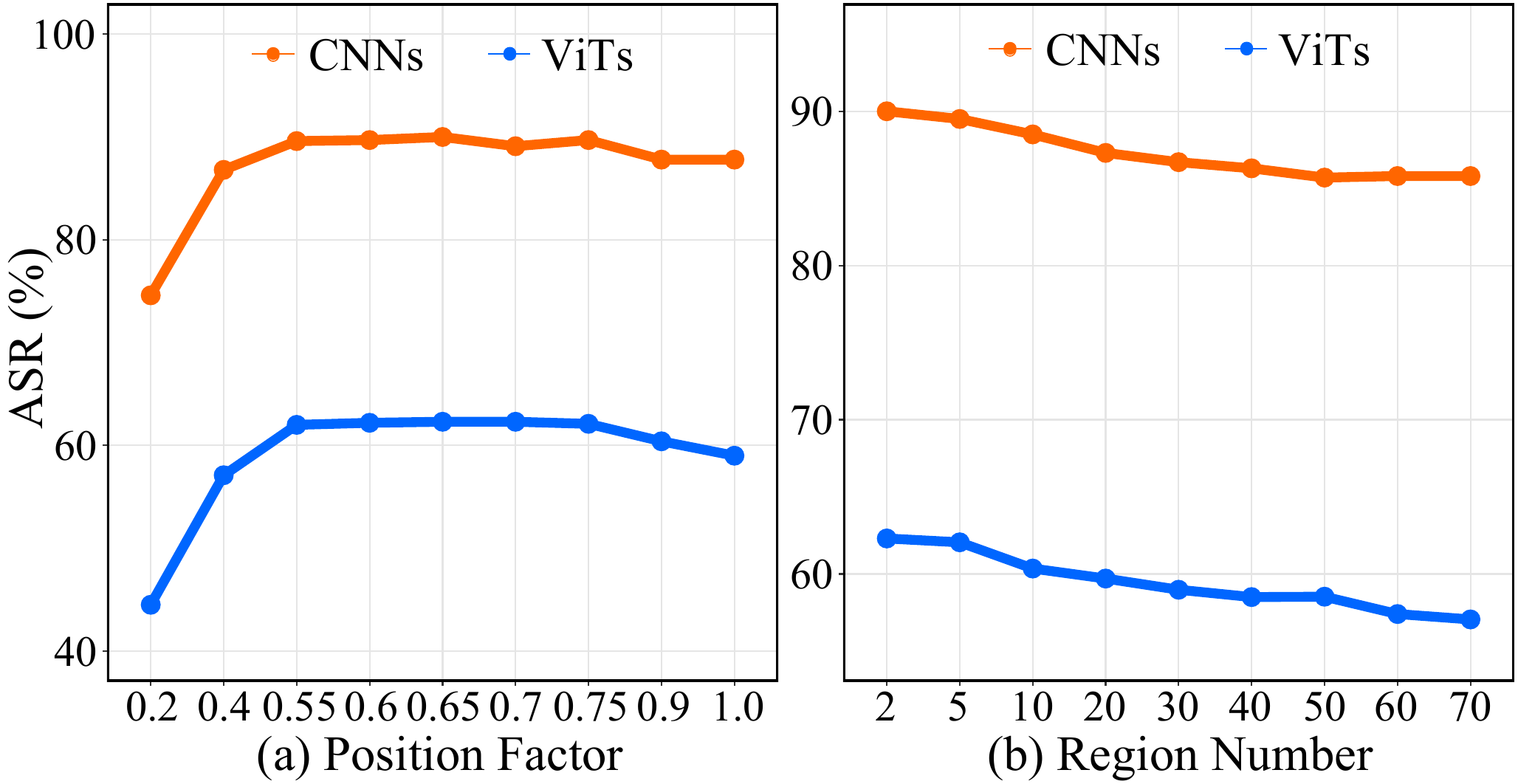}
\caption{Ablation studies of (a) position factor $\lambda$ and (b) region number $N_R$.}
\label{pfrn}
\end{figure}

\textbf{The position factor} $\lambda$ determines the position of the interpolated points in each interval along the straight path.
As can be seen from Figure \ref{pfrn} (a), the attack performance is not sensitive to this $\lambda$ and a moderate $\lambda=0.65$ leads to a satisfactory attack performance.

\textbf{The region number} $N_R$ is a critical parameter in LBQ. A large $N_R$ indicates that the produced baseline is close to the input, whereas a small $N_R$ means that the generated baseline is close to a black image baseline.
Figure \ref{pfrn} (b) demonstrates that transferability diminishes as $N_R$ increases.
% When the produced baseline closely approximates the input, the integration paths shorten, causing interpolated points along each path to become more similar. This similarity leads to greater gradient homogeneity, increasing the risk of overfitting to the surrogate model and consequently reducing transferability. 
This is because a larger $N_R$ makes the produced baseline closely approximate the input.
Therefore, the integration paths become shorter, causing interpolated points along each path to become more similar. This similarity leads to greater gradient homogeneity, increasing the risk of overfitting to the surrogate model and consequently reducing transferability.
Note that $N_R=1$ means the black image baseline.

\textbf{The input transformation} $T$ may also influence the attack performance. 
As can be seen from Table \ref{trans_abl}, removing either of the existing transformations or replacing it with another, common transformation, such as blur or noise, will decrease the transferability.

\begin{table}[!t]
\centering
\begin{tabular}{ccc}
\toprule[1pt]
Transformation & CNNs        & ViTs          \\ \midrule[1pt]
Affine Transformation (AF)             & 82.1        & 53.1          \\
Resizing\&Padding (RP)             & 85.4        & 57.1          \\
AF+Blur        & 83.1        & 51.2          \\
RP+Blur        & 85.6        & 54.9          \\
AF+Noise       & 85.6        & 55.5          \\
RP+Noise       & 87.5        & 58.4          \\
AF+RP (MuMoDIG)         & \textbf{90.0} & \textbf{62.3} \\ \bottomrule[1pt]
\end{tabular}
\caption{Ablation studies of the input transformation $T$.}
\label{trans_abl}
\end{table}

\section{Conclusion and Outlook}

In this paper, we have improved the transferability of integrated gradients (IG)-based attacks by refining their integration paths in three aspects: multiplicity, monotonicity, and diversity.
Concretely, we propose the Multiple Monotonic Diversified Integrated Gradients (MuMoDIG) attack, which can craft highly transferable adversarial examples on various models and defenses.
In particular, the design of MuMoDIG is supported by our theoretical analyses of the fundamental differences in using IG for model interpretation and transferable attacks.
%% save space, remove in the camera-ready version
For future work, we would continue to advance the use of integrated gradients in transferable attacks from the perspectives of baseline generation and more reasonable integration paths. Additionally, we would further promote the fusion of other interpretability methods with transferable attacks for more explainable model evaluations.

\section*{Acknowledgments}
This research is supported by the National Key Research and Development Program of China (2023YFE0209800), the National Natural Science Foundation of China (62406240, T2341003, 62376210, 62161160337, 62132011, U2441240, U21B2018, U20A20177, 62206217), the Shaanxi Province Key Industry Innovation Program (2023-ZDLGY-38).

\bibliography{aaai25}

\clearpage
\appendix

%\toprule[1pt]
%\midrule[1pt]
%\bottomrule[1pt]

\section{Appendix}
\subsection{Proof of Proposition 1}
This section will prove that the non-monotonic path will degrade the attack performance. 
To make it more intuitive, we use BIM as the attack and consider the scenario before overfitting occurs during the iterative process

$Proof.$
Considering a surrogate model $f$, an interpolation coefficient $\alpha$, an adversarial example $x_t$ at the $t$-th iteration and a baseline $b$, according to the Definition 1, a nonmonotonic path is charactirized by
\begin{equation}
    \scriptsize
    \label{nomom}
\text{sign}(\frac{{\partial f({x_t} + \alpha  \cdot ({x_t} - b))}}{{\partial {x_t}}}) \ne \text{sign}(({x_t} - b) \cdot \frac{{\partial f({x_t} + \alpha  \cdot ({x_t} - b))}}{{\partial {x_t}}}),
\end{equation}
where the gradient and the product of the gradient and the path (element-wise multiplication) have inconsistent element symbols. 

Assuming a special case when an image with only one pixel, the update direction of the adversarial example at $t$-th iteration is ${\rm{sign}}(({x_t} - b) \cdot \frac{{\partial f({x_t} + \alpha  \cdot ({x_t} - b))}}{{\partial {x_t}}})$. Since the sign of gradient ${\rm{sign}}(\frac{{\partial f({x_t} + \alpha  \cdot ({x_t} - b))}}{{\partial {x_t}}})$ represents the direction of larger loss, the update direction ${\rm{sign}}(({x_t} - b) \cdot \frac{{\partial f({x_t} + \alpha  \cdot ({x_t} - b))}}{{\partial {x_t}}})$, with completely opposite sign, will lead to the lower loss and thus degrade the attack performance.

In the normal case, although not all the sign of elements are opposite, those opposite parts will also prevent the increase of loss, with the extent depending on the number of opposing elements.

\subsection{Comparison with PAM and NAA}

It is important to note that PAM \cite{pam} is not an IG-based attack, as the value of its "path" does not contribute to the numerical computation of the gradient during the iterative process. While for NAA \cite{naa}, the main difference lies in how IG is applied: NAA uses IG to compute feature weights, while our method integrates IG directly into the iterative process. Additionally, NAA requires selecting specific feature layers, with its original implementation focused only on Inception and ResNet architectures. In contrast, our method is model-agnostic and applicable to various CNNs and ViTs.
\begin{table}[ht]
\centering
\begin{tabular}{lcccc}
\toprule
\textbf{Attack} & \textbf{CNNs} & \textbf{ViTs} & \textbf{AT} & \textbf{NRP} \\
\midrule
PAM      & 76.5 & 36.8 & 46.7 & 56.6 \\
NAA      & 73.2 & 39.5 & 46.1 & 41.4 \\
MuMoDIG  & \textbf{92.5} & \textbf{62.3} & \textbf{51.9} & \textbf{60.7} \\
\bottomrule
\end{tabular}
\caption{Comparison with PAM and NAA. The surrogate model used for all experiments is RN-18, ensuring a fair comparison across methods.}
\label{ap_pamnaa}
\end{table}

We provide a detailed comparison between our method, MuMoDIG, and other concept-related approaches, including PAM  and NAA. The results, as shown in Table \ref{ap_pamnaa}, clearly demonstrate the superiority of our method across different model architectures and defense mechanisms. Specifically, MuMoDIG achieves significantly higher success rates on both CNNs and ViTs compared to PAM and NAA. For instance, MuMoDIG outperforms PAM on CNNs by a margin of 16.0\% and on ViTs by 25.5\%. Similarly, against adversarial training (AT) and neural representation purifiers (NRP), MuMoDIG consistently demonstrates stronger performance, surpassing PAM by 5.2\% and 4.1\%, respectively.

These findings highlight the robustness and effectiveness of MuMoDIG, making it a superior choice for crafting transferable adversarial examples in diverse scenarios.

\end{document}